\documentclass[aps, prl,twocolumn,showpacs,superscriptaddress]{revtex4-1}

\usepackage{braket}

\usepackage{amsfonts}
\usepackage{amsmath}
\usepackage{amssymb}
\usepackage{bm}
\usepackage{dcolumn}
\usepackage{epsfig}
\usepackage{graphicx}
\usepackage{graphics}
\usepackage[latin1]{inputenc}
\usepackage{latexsym}
\usepackage{rotating}
\usepackage[colorlinks=true]{hyperref}
\usepackage[usenames]{color}
\usepackage{yfonts}
\usepackage{xspace} 
\usepackage{mathrsfs}
\usepackage{subfigure}
\usepackage{enumitem}
\usepackage{tabularx}
\usepackage{booktabs}
\usepackage{siunitx}
\usepackage{array}
\usepackage[normalem]{ulem}
\newcolumntype{L}[1]{>{\raggedright\let\newline\\\arraybackslash\hspace{0pt}}m{#1}}
\newcolumntype{C}[1]{>{\centering\let\newline\\\arraybackslash\hspace{0pt}}m{#1}}
\newcolumntype{R}[1]{>{\raggedleft\let\newline\\\arraybackslash\hspace{0pt}}m{#1}}

\usepackage{tabularx}
\usepackage{array}
\usepackage{tabulary}

\newcolumntype{Y}{>{\centering\arraybackslash}X}
\newcolumntype{S}{>{\hsize=.2\hsize}X}
\newcolumntype{M}{>{\centering\arraybackslash}S}

\usepackage{ulem}
\definecolor {darkgreen}{rgb}{0.2,0.7,0.2}
\graphicspath{{../Figures/}}

\newcommand{\be}{\begin{equation}}
\newcommand{\ee}{\end{equation}}
\newcommand\ba{\begin{eqnarray}}
\newcommand\bse{\begin{subequations}}
\newcommand\ea{\end{eqnarray}}
\newcommand\ese{\end{subequations}}

\newcommand{\homg}{{\mbox{\tiny homog}}}
\newcommand{\inhom}{{\mbox{\tiny inhomog}}}
\newcommand{\RL}{{\mbox{\tiny R,L}}}
\newcommand{\Ri}{{\mbox{\tiny R}}}
\newcommand{\Le}{{\mbox{\tiny L}}}
\newcommand{\mat}{{\mbox{\tiny mat}}}

\newcommand{\GR}{{\mbox{\tiny GR}}}

\begin{document}

\title{Gravitational Waves Probes of Parity Violation in Compact Binary Coalescence}

\author{Stephon H. Alexander}
\affiliation{Brown University, Department of Physics, Providence, RI 02912, USA.}
\author{Nicol\'as Yunes}
\affiliation{eXtreme Gravity Institute, Department of Physics, Montana State University, Bozeman, MT 59717, USA.}

\date{\today }

\begin{abstract}
%
Is gravity parity violating? Given the recent observations of gravitational waves from coalescing compact binaries, we develop a strategy to find an answer with current and future detectors.  We identify the key signatures of parity violation in gravitational waves: amplitude birefringence in their propagation and a modified chirping rate in their generation. We then determine the optimal binaries to test the existence of parity violation in gravity, and prioritize the research in modeling that will be required to carry out such tests before detectors reach their design sensitivity.
\end{abstract}
\keywords{cosmology}
\pacs{}
\maketitle

\emph{Introduction}. Einstein's theory of General Relativity (GR) has made striking predictions in the weak field of the Solar System~\cite{Will:2014kxa}, and in the strong field of binary pulsars~\cite{Stairs2003}. The groundbreaking detections of gravitational waves (GWs) by the LIGO-Virgo scientific collaboration~\cite{Abbott:2016blz} have now led to the first confirmations of Einstein's theory~\cite{TheLIGOScientific:2016src,TheLIGOScientific:2016pea,Monitor:2017mdv} in \emph{extreme gravity}~\cite{Yunes2013}, where gravity is strong and dynamical. GW probes of extreme gravity go beyond confirmations of Einstein's theory through theoretical physics implications~\cite{Yunes:2016jcc}. For example, we have confirmed that the (real part of the) group velocity of gravity is frequency independent~\cite{Mirshekari:2011yq,TheLIGOScientific:2016src,Yunes:2016jcc}, constraining quantum-gravity-inspired theories~\cite{Will:1997bb,Stavridis:2010zz,TheLIGOScientific:2016src}. We have also confirmed that GWs carry energy in a predominantly quadrupolar way, constraining theories that predict dipole radiation~\cite{Yagi:2011xp,Yunes:2011we,Yagi:2015oca,Yunes:2016jcc}. The recent observation of a neutron star (NS) merger confirmed that the (frequency-independent part of the) speed of gravity is equal to that of light~\cite{Blas:2016qmn,Cornish:2017jml,Monitor:2017mdv}, constraining dark energy models~\cite{Ezquiaga:2017ekz,Sakstein:2017xjx,Creminelli:2017sry,Boran:2017rdn,Baker:2017hug}. 

One important aspect that has not received much attention yet is gravitational parity invariance in extreme gravity~\cite{Alexander:2007kv,Yunes:2008bu,Yunes:2010yf,Alexander:2009tp}. Analyzing parity in GR is subtle because of its inherent coordinate covariance, but we can think of it as a symmetry of the Lagrangian under the parity operator in a suitable $3+1$ decomposition. GR is thus a parity-invariant theory.  In fact, the weak force is the only interaction that maximally violates parity, which explains the observed decay of nuclei and mesons~\footnote{To preserve unitarity in quantum theories with anomalous gauge currents, parity-violating, anomaly-cancelling terms must be included in the action, such as the triangle and gravitational anomalies~\cite{Bell:1969ts,1969PhRv..177.2426A}, and their 10-dimensional generalizations in heterotic string theory~\cite{Green:1987sp}.} (e.g.~$\beta$-decay of cobalt-60~\cite{1957PhRv..105.1413W} and neutral kaon decay~\cite{1956PhRv..104..254L}).  Moreover, parity-violation in GR plays a key role in cosmological baryogenesis and relates the baryon asymmetry index to CMB observables~\cite{Alexander:2004xd,Alexander:2009tp}. 

Given the possibility of gravitational parity violation in nature, one is urged to search for observational signatures in extreme gravity. This note studies the theory of \emph{generic} gravitational parity violation, showing that it \emph{reduces} to dynamical Chern-Simons gravity~\cite{Jackiw:2003pm,Alexander:2009tp}). We then summarize the key signatures of generic parity violation in GW observables, classifying them into generation and propagation effects. We use these signatures to identify the best compact binaries to detect or constrain parity violation with current and future GW interferometers. The binaries identified allow us to discuss the research in GW modeling that ought to be prioritized to carry out such tests in the future.

\emph{Parity Violation in Extreme Gravity}. Let us begin by defining parity violations more precisely.  Consider a spatial hypersurface (a surface of constant time) on which we define the parity operator $\hat{P}$ as the spatial reflection of the spatial triad $e^{I}{}_{i}$ that defines the spatial coordinate system, i.e.~$\hat{P} \left[e^{I}{}_{i}\right] = - e^{I}{}_{i}$, where Latin letters are spatial indices~\cite{Misner:1973cw}. The line element is parity invariant because it denotes the differential length of a four-dimensional interval. Performing a 3+1 decomposition of the line element, 
\be
ds^{2} = -\alpha^{2} dt^{2} + h_{ij} \left(dx^{i} + \beta^{i} dt\right) \left(dx^{j} + \beta^{j} dt\right)\,,
\ee
where $G = 1 = c$~\cite{Misner:1973cw}, we see that its parity invariance implies the lapse $\alpha$ and the spatial metric $h_{ij}$ must both be even, while the shift $\beta^{i}$ must be odd, and so the extrinsic curvature $K_{ij}$ is even. The Ricci scalar    
\be
R = \;^{3}\!R + K^{ij} K_{ij} - K^{2} - 2 \nabla_{\alpha}\left(n^{\beta} \nabla_{\beta} n^{\alpha} - n^{\alpha} \nabla_{\beta}  n^{\beta}\right)\,,
\ee
where Greek letters are spacetime indices~\cite{Misner:1973cw}, must then be even because both the induced Ricci scalar $^{3}\!R$ and the normal vector $n^{\alpha}$ are even. 

How can one construct a parity violating interaction, a pseudo-scalar, to add to the Lagrangian? Consider first using only the curvature tensor~\cite{Yagi:2015oca}. Because of the symmetries of the Riemann tensor and the Bianchi identity $R_{\mu [\nu \alpha \beta]} = 0$, there are no pseudo-scalars at linear order in curvature. At second order, the only possibility is the Pontryagin density
\be
^{*}\!RR := {}^{*}\!R^{\alpha \beta \delta \gamma} R_{\alpha \beta \delta \gamma}\,,
\ee
where ${}^{*}\!R^{\alpha \beta \delta \gamma}$ is the dual Riemann tensor. This quantity, however, is a topological invariant, so it can locally be written as the divergence of a four-current~\cite{Jackiw:2003pm,Alexander:2009tp}, which does not contribute to the field equations. The simplest way for this term to contribute is to multiply it by a function of a field. 

The simplest effective action that incorporates gravitational parity violation (to second-order in the curvature and with a single scalar field) is one that adds to the Einstein-Hilbert Lagrangian a dynamical and anomalous current $J_{5}^{\mu}$ sourced by the Pontryagin density. The field equations for this simplest gravitational parity-violating effective theory are  
\begin{align}
G_{\mu \nu} + \frac{\alpha}{\kappa_{g}} C_{\mu \nu} &= \frac{1}{2 \kappa} \left(T_{\mu \nu}^{\mat} + T_{\mu \nu}^{(J_{5})}\right)\,,
\\
\label{eq:scalar-EOM}
\nabla_{\mu} J_{5}^{\mu} &= - \alpha \; ^{*}\!RR\,,
\end{align}
where $G_{\mu \nu}$ is the Einstein tensor, $T_{\mu \nu}^{\mat}$ is the matter stress-energy tensor, $\alpha$ is a coupling constant (with units of length squared), $\kappa_{g} = (16 \pi G)^{-1}$, $C_{\mu \nu}$ is an interaction term that depends on $J^{\mu}_{5}$ and its derivative as well as on the Ricci and Riemann tensors, and $T_{\mu \nu}^{(J_{5})}$ is the stress-energy tensor for the anomalous current~\cite{Alexander:2009tp}. The equation of motion for the anomalous current, Eq.~\eqref{eq:scalar-EOM}, is \emph{identical} to an anomaly-cancelling term in chiral gauge theories. Therefore, this theory is a good toy model for phenomenological studies of gravitational parity violation. The above theory reduces to dynamical Chern-Simons gravity~\cite{Jackiw:2003pm,Alexander:2009tp} when one identifies $\nabla^{\mu} \vartheta \to J_{5}^{\mu}$ for a dynamical pseudo-scalar field $\vartheta$. To make contact with previous results, we will here choose the $\vartheta$ parametrization of the parity-violating effect, but in spite of its similarities with dynamical Chern-Simons theory, one should remember that the effects we consider here are generic.

\emph{Parity-Violating Propagation Effects}. The main propagation effect is \emph{polarization mixing}, i.e.~the initial polarization state is not conserved under propagation. The field-theoretic way to understand such an effect is through a modification of the propagator~\cite{Yunes:2010yf}, but a more familiar approach is to modify the \emph{dispersion relation}. The latter is obtained by linearizing the field equations about a background, like the Friedmann-Robertson-Walker (FRW) spacetime, with a wave-like perturbation 
\be
\label{eq:wave-ansatz}
h_{\mu \nu} = A_{\mu \nu} e^{-i [\phi(t) - k_{i} x^{i}]}\,,
\ee
where $\phi(t)$ and $k^{i}$ are the wave's time-dependent phase and wave vector, and $A_{\mu \nu}$ is the polarization-dependent amplitude. Gravitational parity violation leads to a \emph{purely-imaginary}, polarization-dependent modification to the dispersion relation
\be
i \ddot{\phi} + 3 i H \dot{\phi} + \dot{\phi}^{2} - k_{i} k^{i} = i \lambda_{\RL} \; \dot{\phi} \; g(\dot{\vartheta},\ddot{\vartheta})\,,
\ee
where $H$ is the Hubble parameter, $g(\cdot)$ encodes parity violation and $\lambda_{\RL} = \pm 1$ for right/left polarizations. 

We can compare this dispersion relation to dark energy emulators in modified gravity~\cite{Ezquiaga:2017ekz}  
\be
i \ddot{\phi} + (3 + \alpha_{M}) i H \dot{\phi} + \dot{\phi}^{2} - \left(1 + \alpha_{T}\right) k_{i} k^{i} = 0\,,
\ee
where $\alpha_{T}$ determines the speed of tensor modes, $c_{g}^{2} = 1 + \alpha_{T}$, and $\alpha_{M}$ is related to the running of the Plank mass. Comparing these equations, $\alpha_{T} = 0$ and 
\be
\alpha_{M}  =   \lambda_{\RL} \; H^{-1} \; g(\dot{\vartheta},\ddot{\vartheta})\,.
\ee
\emph{Since the recent coincident observation of gamma rays and GWs emitted in a NS merger~\cite{Monitor:2017mdv} only constrains $\alpha_{T}$, it places no bounds on gravitational parity violation.}

Why does such a modification encode parity violation? The modification has different signs depending on whether the wave is right- or left-polarized, forcing a different evolution equation for the different polarization states. Solving the dispersion relation by linearizing about the GR solution $\bar{\phi}$, i.e.~$\phi = \bar{\phi} + \lambda_{\RL} \delta \phi$ with $\delta \phi \ll \bar{\phi}$ small, we find
\be
\delta \phi = \frac{1}{2}  \; i \int  g(\dot{\vartheta},\ddot{\vartheta}) dt\,,
\ee
where we assume $\delta \ddot{\phi} \ll \dot{\bar{\phi}} \; \delta \dot{\phi}$. Reinserting this solution into Eq.~\eqref{eq:wave-ansatz}, we find
\be
\label{eq:rl-bire}
h_{\RL} = h_{\RL}^{\GR} e^{-i \lambda_{\RL} \delta \phi} \sim h_{\RL}^{\GR} \left[1 + \frac{1}{2}  \lambda_{\RL} \int  g(\dot{\vartheta},\ddot{\vartheta}) dt \right]\,,
\ee
where we projected the metric perturbation into a left/right basis. 

The effect of gravitational parity violation is an enhancement/suppression of the right/left-polarized content of a GW, an \emph{amplitude birefringence}. We can see this more clearly by mapping from the left-/right-polarization basis to a linear $(+,\times)$ basis, using $h_{+} = (h_{\Ri} + h_{\Le})/\sqrt{2}$ and $h_{\times} = i(h_{\Ri} - h_{\Le})/\sqrt{2}$. Doing so, we find~\footnote{Comparing to~\cite{Alexander:2007kv}, $\delta \phi = i \, v$ in their notation.}
\begin{align}
h_{+} &= \bar{h}_{+} - \delta \phi \; \bar{h}_{\times}\,,
\qquad
h_{\times} = \bar{h}_{\times} + \delta \phi \; \bar{h}_{+}\,.
\end{align}
where as before $\bar{h}_{\pm}$ are the $(+,\times)$ polarizations in GR. Observe a mixing of the $(+,\times)$ polarization that is enhanced upon propagation. 

For concreteness, specialize these generic considerations to the dynamical Chern-Simons case. This calculation was first done in~\cite{Alexander:2007kv,Yunes:2008bu,Yunes:2010yf}, who found that the dispersion relation takes the form of Eq.~\eqref{eq:wave-ansatz} with~\footnote{This expression is different from that in~\cite{Alexander:2007kv,Yunes:2008bu,Yunes:2010yf} because the Pontryagin interaction term is here defined as $4 \alpha/\kappa_{g}$ times that of~\cite{Alexander:2007kv,Yunes:2008bu,Yunes:2010yf}. Also, this result was found originally in the non-dynamical version of this theory, which lacks a stress-energy contribution from the $\vartheta$ field; we have found that the same dispersion relation is obtained in the dynamical theory when $\vartheta$ is only time-dependent.} 
\be
g(\dot{\vartheta},\ddot{\vartheta}) = -4 (\alpha/\kappa_{g}) \, k \, \left( \ddot{\vartheta} - H \dot{\vartheta}\right)\,,
\ee
where $k = |k_{i}k^{i}|$, and the phase correction is 
\be
\label{eq:delta-phi-dCS}
\delta \phi = -2 i (\alpha/\kappa_{g})  \int k(t) \left(\ddot{\vartheta} - H \dot{\vartheta}\right) dt\,,
\ee
which is imaginary and proportional to the wave frequency and distance traveled. 

The dephasing can be further specified by solving for the evolution of the scalar field. In dynamical Chern-Simons gravity, this evolution is controlled by Eq.~\eqref{eq:scalar-EOM}, whose solution has a homogeneous and an inhomogeneous piece. In a FRW background, the former implies $\ddot{\vartheta} = -3 H \dot{\vartheta}$, and thus~\cite{Yunes:2010yf}
\begin{align}
\label{eq:deltaphiH0}
\delta \phi_{\homg} &= -8 (\alpha/\kappa_{g})  i  \int k(t) \,  H(t) \, \dot{\vartheta} \, dt\,.
\\
\label{eq:deltaphiH}
&= -8 (\alpha/\kappa_{g}) i \; \omega_{0} \dot{\vartheta}_{0} \, z \,, 
\end{align}
expanding in small redshift $z \ll 1$ and assuming a monochromatic wave with angular frequency $\omega(t) = \omega_{0}$.  

One might worry that this effect is vanishingly small today, if $\vartheta$ were produced in the early Universe and then Hubble diluted upon evolution. The $\vartheta$ field does obey $\ddot{\vartheta} = -3 H \dot{\vartheta}$, forcing it to exponentially decay with time. Any value for $\dot\vartheta$ set by cosmological boundary conditions, at the beginning of radiation domination, is exponentially suppressed today. The $\vartheta$ field, however, is constantly being regenerated by $^{*}\!RR$ sources due to the dynamics of compact binary mergers, most of which occur at redshifts $z < 5$~\cite{Cutler:2002me}. If these mergers regenerate $\dot{\vartheta}$ at $z<0.1$, there is not enough evolution to force $\dot{\vartheta}$ to decay to zero by today.  

The inhomogeneous solution is more difficult to calculate, but we can show its effects are subdominant. The solution to Eq.~\eqref{eq:scalar-EOM} in the small-coupling, weak-field approximation was found in~\cite{Yagi:2011xp}. Using this solution non-perturbatively in Eq.~\eqref{eq:delta-phi-dCS}, neglecting the Hubble-dependent term and integrating over the wave's travel time, while assuming the orbital frequency is approximately constant~\footnote{This frequency is not constant in an inspiral, but its variation occurs on a radiation-reaction timescale, so it is suppressed by ${\cal{O}}(c^{-5})$.}, we find $\delta \phi \sim (m \omega)^{13/3} = v^{13}$, with $m$ the total mass of the binary, $\omega$ its angular frequency and $v$ its orbital velocity. This effect is of very high post-Newtonian (PN) order~\footnote{A PN expansion is one in powers of $v/c$~\cite{Blanchet:2006zz}; a term proportional to $(v/c)^{2N}$ relative to its leading-order expression is said to be of Nth PN order. Therefore, this effect is of $5.5$ PN order in the amplitude.} and subdominant relative to the the homogeneous solution. 

\emph{Parity-Violating Generation Effects}. The main generation effect is a \emph{modified energy loss, inspiral rate and chirping rate}. The field-theoretic way to understand this is to realize that parity-violation requires a parity-violating current that satisfies an anomalous conservation equation. When the right-hand side of Eq.~\eqref{eq:scalar-EOM} is evaluated for inspiraling binaries, the scalar field becomes wave-like, carrying energy-momentum away as it propagates, draining the orbital binding energy and forcing the system to decay faster, as first found in~\cite{Yagi:2011xp}. This acceleration in the rate of decay affects the \emph{chirping rate}, i.e.~the rate at which the orbital and the GW frequency increases with time, which is encoded in the GW observable.

Let us now specialize these generic considerations to dynamical Chern-Simons gravity. This calculation was first done in~\cite{Yunes:2010yf}, who found that the rate at which $\vartheta$ carries energy away from a binary is
\be
\label{eq:Edot-def}
\dot{E}^{(\vartheta)} = \int_{S_{\infty}} \left< \dot{\vartheta}^{2} \right> r^{2} d\Omega\,,
\ee
where $S_{\infty}$ a 2-sphere at spatial infinity, and the angle-brackets stand for averaging over several wavelengths. 

This energy loss rate can be further specified by prescribing the evolution of the scalar field. As in the propagation case, $\vartheta$ has a homogeneous and an inhomogeneous solution. Assuming a Minkowski background, compatible with the spacetime of a compact binary in the far-zone, the homogeneous solution is 
\be
\vartheta_{\homg} = \sqrt{\alpha} \; \vartheta_{0} \frac{\cos{\Omega (t -r)}}{r} + \sqrt{\alpha} \; \frac{\dot{\vartheta}_{0}}{\Omega} \; \frac{\sin{\Omega (t -r)}}{r}\,,
\ee
where $\Omega$ is its angular frequency, and $(\vartheta_{0},\dot{\vartheta}_{0})$ are the initial value and velocity of the dimensionless field~\footnote{We have pulled out a factor of $\alpha^{1/2}$ for dimensional consistency, since $\vartheta$ is dimensionless here.}. The rate at which a homogeneous $\vartheta$ carries energy away from a binary is
\be
\label{eq:EdotH}
\dot{E}^{(\vartheta)}_{\homg} = \frac{\Omega^{2}}{2}  \alpha  \vartheta_{0}^{2} + \frac{1}{2}  \alpha  \dot{\vartheta}_{0}^{2} \,.
\ee
Unlike in the propagation case, this modification is only active during the generation process, which occurs on a radiation-reaction time scale. The latter is much smaller than the GW time of flight, roughly of ${\cal{O}}(H_{0})$ smaller, making its impact on the evolution of the GW phase negligible.   

Consider now the contribution to the energy loss from the inhomogeneous solution. This calculation was first done in~\cite{Yunes:2010yf,Yagi:2012vf} for binary black holes (BHs), who found that 
\be
\label{eq:EdotIH}
\dot{E}^{(\vartheta)}_{\inhom} = \frac{5}{48} \zeta \; \eta^{2} \, \left(\Delta^{i} \Delta_{i}\right) \, \left(m \omega\right)^{14/3}\,,  
\ee
with $\eta = m_{1} m_{2}/m^{2}$ the symmetric mass ratio, $\zeta \equiv \alpha^{2}/(m_{1}^{2} m_{2}^{2} \kappa_{g})$ a dimensionless coupling constant~\footnote{The dimensionless coupling parameter in dynamical Chern-Simons gravity is sometimes also defined as $\zeta \equiv 16 \pi \alpha^{2}/(m_{1}+m_{2})^{4}$, which is different from the normalization chosen here.}, and $\Delta^{i} \equiv ({m_{2}}/{m}) \chi_{1} \hat{S}_{1}^{i} - ({m_{1}}/{m}) \chi_{2} \hat{S}_{2}^{i}$, with $\chi_{A}$ and $\hat{S}_{A}^{i}$ the dimensionless spin parameter and the spin angular momentum direction of the Ath object. 

Such a modification to the decay rate translates into a correction to the chirping rate. The correction to the GW phase in the Fourier domain is of 2PN order~\cite{Yagi:2012vf}
\be
\label{eq:deltaphi}
\delta \Psi(f) = -\frac{30}{128} \left(\pi {\cal{M}} f\right)^{-5/3} \, \delta C \; \left(\pi \, m \, f\right)^{4/3}\,, 
\ee
where ${\cal{M}} \equiv \eta^{3/5} m$ is the chirp mass, $f$ is the GW frequency, and $\delta C$ is proportional to $\zeta$ and a function of the component masses and the dimensionless spins~\footnote{This correction is not only generated by a modification to the energy flux, but also by modifications to the quadrupole moment of the BHs, and a scalar-dipole interaction, all at 2PN order~\cite{Yagi:2012vf}.}. The above result seems specific to dynamical Chern-Simons gravity, but in reality, it is not, since all it requires is for there to exist a dynamical anomalous current, such as that given by the gradient of a scalar field, whose conservation is parity-violating and anomalous in the sense of Eq.~\eqref{eq:scalar-EOM}. 

\emph{Detection Prospects}. Let us first consider gravitational parity violation in GW \emph{propagation}. The main modification is in Eqs.~\eqref{eq:deltaphiH0} and~\eqref{eq:deltaphiH}, which grows with luminosity distance and GW frequency. Naively, the best sources are late-inspiral/merger events at high redshift, such as supermassive BH mergers, but the correction must not be degenerate with other parameters in the model. The propagation effect is partially degenerate with the inclination angle and the location of the source in the sky~\footnote{Partial degeneracies are also due to higher PN order corrections and spin effects. These, however, depend on frequency and on the binary parameters differently than the effect considered here.}, as first pointed out in~\cite{Alexander:2007kv}. These degeneracies can be broken with a coincident short gamma-ray burst observation~\cite{Monitor:2017mdv}. Therefore, the ideal source for constraints on parity violation in the propagation of GWs are mixed NS/BH mergers with a short gamma-ray burst counterpart, since they enhance the GW propagation time and break the inclination angle degeneracy. 

Consider now constraints on the effect of gravitational parity violation in GW \emph{generation}. The main modification is in Eq.~\eqref{eq:deltaphi}, which depends on $\delta C$, and thus on the BH spins. There is a strong degeneracy between this parity violation modification and the spins, which actually prevented any constraints with the first LIGO observations~\cite{Yunes:2016jcc,Yagi:2017zhb}. The degeneracy, however, can be broken if both spin magnitudes are extracted, which requires either the observation of a spin-precessing BH binary, or the observation of a spinning BH/NS binary. In the former, the amplitude modulations introduced by spin-precession break the degeneracy, while in the latter only the BH spin matters and can be extracted from the spin-orbit modification to the GW phase. NSs are expected to have small dimensionless spins, thus suppressing gravitational parity violation effects in the generation of GWs. 

The effect of gravitational parity violation on GW generation is further enhanced in eccentric binaries. Eccentricity changes the GW phase in GR, making the leading-order term in the Fourier phase a factor of $v^{-19/3}$ larger than in the quasi-circular case for small eccentricities~\cite{Yunes:2009yz}. Eccentricity-dependent corrections to Eq.~\eqref{eq:deltaphi} will enter below $-2$ PN order instead of at $+2$ PN order. Such ``negative'' PN order corrections lead to an enhancement in our ability to detect or constrain them. Therefore, the ultimate source for constraints on gravitational parity violation effects in GW generation are mildly eccentric compact binaries composed either of a NS and a spinning BH or two spinning BHs, in both cases with large spins.
  
With these ideal systems identified, let us now provide a quantitative estimate of the accuracy to which these parity-violating effects could be constrained with future observations. Assuming degeneracies are broken, a single GW detection consistent GR up to statistical uncertainties would imply that (parameterized post-Einsteinian~\cite{Yunes:2009ke}) deformations must be smaller than approximately one over the signal-to-noise ratio  at the dominant GW frequency in detector's sensitivity the band~\cite{Cornish:2011ys,Sampson:2013lpa,Sampson:2014qqa,Yunes:2016jcc}. For a propagation effect, this leads to
\be
\frac{\alpha}{\kappa_{g}} J_{5}^{0} < \frac{1}{8 \rho} (2 \pi F_{0} z)^{-1} \approx 400 \; {\rm{km}}  \,,   
\ee
for a dominant orbital frequency $F_{0}$ of $50$ Hz, which is where ground-based detectors have best sensitivity, and $z=0.1$ with a signal-to-noise ratio of $\rho = 10$. Similarly, constraints on parity violation during GW generation should approximately be
\be
\frac{\sqrt{\alpha}}{\kappa_{g}^{1/4}}   < \frac{15}{\rho^{1/4}} \frac{(\pi m f)^{-1/3}  \eta^{7/10}}{(61969 - 231808 \eta)^{1/4}} \; \frac{m}{\chi_{s}^{1/2} } \approx 33 \; {\rm{km}}
\ee
for a binary with spins co-aligned with the orbital angular momentum. In the last equality, we evaluated the constraint for an equal-mass binary with $m = 10 M_{\odot}$, symmetric dimensionless spin $\chi_{s} = (\chi_{1} + \chi_{2})/2 = 0.5$, dominant GW frequency $f = 100$ Hz and $\rho = 10$. As expected from a Fisher analysis, the constraints are inversely proportional to $\rho$, becoming stronger with redshift in the propagation case and with curvature scale (proportional to $1/m^{2}$) in the generation case. These quantitative estimates are consistent with projected bounds studied in the specific case of dynamical Chern-Simons gravity~\cite{Alexander:2007kv,Yunes:2008bu,Yunes:2010yf,Yagi:2012vf,Yagi:2017zhb}.  

How likely are we to observe the ideal sources discussed above with future observing runs and detectors? During the third observing run, advanced LIGO and Virgo should operate at higher sensitivities (hopefully better by a factor of 2), thus increasing the range the instruments can see by roughly that factor (and the volume by a factor of 8). Even if such events are not observed in the third observing run, they are very likely to be observed when the instruments reach their design sensitivity by 2020, or when improvements are implemented and third-generation detectors are constructed in the next decade. Recent studies suggest that our ability to test GR will improve by somewhere between 5 and 10 orders of magnitude with third-generation detectors, depending on the specific test considered~\cite{Barausse:2016eii,Chamberlain:2017fjl}.  
 
The search for gravitational parity violation described here requires accurate models for the GWs emitted in parity-violating theories. In particular, given the ideal sources discussed above, one needs models that describe GWs emitted by eccentric and spin-precessing NS/BH and BH/BH binaries. The development of analytic models for such GWs is in its infancy even within GR. Analytic models for the GWs emitted by spin-precessing systems~\cite{Chatziioannou:2016ezg,Chatziioannou:2017tdw} and for eccentric models~\cite{Yunes:2009yz,Loutrel:2016cdw,Tanay:2016zog,Boetzel:2017zza} have only recently become available in GR. Their generalization to include gravitational parity violation requires the re-calculation of the solution to the Kepler problem and to the spin-precession equations in the inspiral phase, which should be considered a priority. 

\acknowledgements N.~Y.~acknowledges support from NSF CAREER grant PHY-1250636 and NASA grants NNX16AB98G and 80NSSC17M0041.

\bibliography{refs}{}
\bibliographystyle{ieeetr}

\end{document}